# Intrinsic valley Hall transport in atomically thin MoS$_2$

Zefei Wu[1*]†, Benjamin T. Zhou[1]†, Gui-Bin Liu[3], Jiangxiazi Lin[1], Tianyi Han[1], Liheng An[1], Yuanwei Wang[1], Shuigang Xu[1], Gen Long[1], Chun Cheng[4], Kam Tuen Law[1], Fan Zhang[2*], Ning Wang[1*]

[1]Department of Physics and the Center for Quantum Materials, the Hong Kong University of Science and Technology, Hong Kong, China.

[2]Department of Physics, University of Texas at Dallas, Richardson, TX 75080, USA.

[3]School of Physics, Beijing Institute of Technology, Beijing 100081, China.

[4]Department of Materials Science and, Southern University of Science and Technology, Shenzhen 518055, China.

*Correspondence to: phwang@ust.hk (N.W.); zhang@utdallas.edu (F.Z.); wzefei@connect.ust.hk (Z.W.)

†These authors contributed to this work equally.

**Electrons hopping in two-dimensional honeycomb lattices possess a valley degree of freedom in addition to charge and spin. In the absence of inversion symmetry, these systems were predicted to exhibit opposite Hall effects for electrons from different valleys. Such valley Hall effects have been achieved only by extrinsic means, such as substrate coupling, dual gating, and light illuminating. Here, we report the first observation of intrinsic valley Hall transport without any extrinsic symmetry breaking in the non-centrosymmetric monolayer and trilayer $MoS_2$, evidenced by considerable nonlocal resistance that scales cubically with local resistance. Such a hallmark survives even at room temperature with a valley diffusion length at micron scale. By contrast, no valley Hall signal is observed in the centrosymmetric bilayer $MoS_2$. Our work elucidates the topological quantum origin of valley Hall effects and marks a significant step towards the purely electrical control of valley degree of freedom in topological valleytronics.**

Electron valley degree of freedom emerges as local extrema in the electronic band structures. Inequivalent valleys, well separated in the Brillouin zone, can be energetically degenerate due to symmetry and serve as novel information carriers controllable via external fields[1-4]. A feasible means to manipulate such a valley degree of freedom is through a valley Hall effect (VHE)[3-7]. Analogous to an ordinary Hall effect, in which a transverse charge current is driven by a uniform magnetic field in real space, a transverse valley current in the VHE is produced by valley-contrasting Berry curvatures in momentum space. Upon the application of an external electric field, the curvatures drive carriers from different valleys to traverse in opposite directions. Therefore, the VHE has been a major theme in the study of valleytronics, particularly in those 2D materials featuring K and K' valleys in their hexagonal Brillouin zones[8-17].

As Berry curvature is even under spatial inversion ($P$) and odd under time reversal ($T$), the VHE cannot survive when both $P$ and $T$ symmetries are present. To achieve VHEs in monolayer and bilayer graphene, an elaborately aligned h-BN substrate[8] and a strong dual gating field[9,10] were respectively utilized to break the $P$ symmetry. To excite VHEs in specific valleys[15,16], circularly polarized lights[18-20] were used for breaking the $T$ symmetry in atomically thin transition-metal dichalcogenides (TMDC). Monolayer TMDCs have direct band gaps of optical frequencies at two inequivalent K-valleys[21,22], due to the intrinsic $P$ asymmetry in their unit cells depicted in Fig. 1a. Thus, Berry curvatures with opposite signs naturally emerge at the two K-valleys. Moreover, a mirror symmetry locks the spin and valley indices of the sub-bands split by the spin-orbit couplings, both of which are flipped under $T$; the spin conservation suppresses the inter-valley scattering. Therefore, monolayer TMDCs have been deemed an ideal platform for realizing intrinsic VHE without extrinsic symmetry breaking[13,14].

However, the quantum transport in atomically thin TMDCs has been a long-standing challenge due to the low carrier mobility and the large contact resistance in their field-effect devices prepared by the exfoliation method. Recent breakthroughs in the fabrication of low-temperature ohmic contacts for high-mobility 2D TMDC devices[23-26] have already facilitated the observation of transport hallmarks of Q-valley electrons[26,27], K-valley holes[28-30], and Γ-valley holes[31]. These discoveries have revealed the rich and unique valley physics in the platform of atomically thin TMDCs.

In this work, we systematically examine *intrinsic* VHEs through measuring nonlocal resistances and their layer dependence in n-type 2H-MoS$_2$. For the first time, we observe nonlocal resistances that exhibit cubic power-law scaling with the local resistances in the monolayers and trilayers, evidencing intrinsic VHEs. Because of the large intrinsic bandgaps of TMDCs, such VHEs can even be observed at room temperature in our monolayer devices. Beyond critical carrier densities ($5\times10^{10}$ cm$^{-2}$ for monolayers and $1.1\times10^{11}$ cm$^{-2}$ for trilayers), the cubic scaling turns into linear scaling. Notably, only linear scaling is observed in bilayer MoS$_2$, where the $P$ symmetry is restored. Intriguingly, although the monolayer and trilayer feature respectively K- and Q-valleys near their conduction-band edges, they display comparable valence-band Berry curvatures, valley Hall signatures, and micron-sized valley diffusion lengths. Our results not only offer the first experimental evidence for the intrinsic VHE but also help elucidate its fundamental origin in odd-layer TMDCs and pave the way for realizing room-temperature low-dissipation valleytronics by purely electronic means.

Figure 1b is an optical image of a monolayer MoS$_2$ field-effect transistor. The structure of this transistors is sketched in Fig. 1c. The device fabrication process includes a dry transfer step followed by a selective etching step[25,26,31] (Please see Methods and Supplementary Fig.1 for details). A low contact barrier formed on the n-type MoS$_2$ is evidenced by the *I-V* curves and the field-effect mobilities $\mu$ varied from 1,000 cm$^2$V$^{-1}$s$^{-1}$ to 25,000 cm$^2$V$^{-1}$s$^{-1}$ at $T = 2$ K (Supplementary Figs. 2 and 3).

As for the electronic measurement, an inverse VHE is exploited to detect a valley current, as sketched in Fig. 1d. An applied current $I_{12}$ through probes 1 and 2 induces charge imbalance in a remote region, as measured by the voltage drop $V_{34}$ between probes 3 and 4 (Supplementary Fig. 4). The nonlocal resistance $R_{NL} = V_{34} / I_{12}$ mediated by the valley Hall current was predicted[32] to present cubic power-law dependence on the local resistance $R_L = V_{24} / I_{13}$.

**Nonlocal transport in monolayer MoS$_2$.**

Nonlocal resistance $R_{NL}$ in an n-type monolayer MoS$_2$ (sample B of length $L = 6$ μm and width $W = 1.5$ μm illustrated in Fig. 1d), measured as a function of gate voltage $V_g$ at varied temperatures, is shown in Fig. 1e. A giant $R_{NL}$ is observed in the range of $V_g \sim -15$ to $-25$ V that amounts to the electron density n $\sim 10^{10}$ to $10^{11}$ cm$^{-2}$. In particular, the observed $R_{NL} \sim 10^6$ Ω exceeds the classical ohmic contribution $R_{CL} = R_L \frac{W}{\pi L} e^{-\pi L/W} \sim 10^4$ Ω by two orders of magnitude in the range of $V_g \sim -15$ to $-18$ V at 2 K and $V_g \sim -22$ to $-25$ V at 300 K. Another unexpected feature of $R_{NL}$ is its $V_g$ dependence. In sharp contrast to the classical contribution $R_{CL}$, which decreases gradually with increasing $V_g$, the observed $R_{NL}$ drops by at least one order of magnitude within an increase of several volts in $V_g$. Both the pronounced nonlocal signal and its unusual sensitivity to $V_g$ suggest that the observed $R_{NL}$ has a physical origin different from the classical ohmic contribution $R_{CL}$.

The temperature dependence of $R_L$ and $R_{NL}$ uncovers the mesoscopic mechanism of both the local and nonlocal transport. The conduction can be separated into three regimes:

the thermal activation (TA) at 250 K<T<130 K, the nearest-neighbor hopping (NNH) at 130 K<T<60 K, and the variable-range hopping (VRH) below 60 K (sample A of $L = 3.6$ μm and $W = 1.5$ μm, see Fig. 2f, Supplementary Figs. 5a and 5b). Interestingly, the characteristic temperatures of both NNH and VRH for $R_{NL}$ are much larger than those for $R_L$ in the range of $V_g \sim -60$ to $-58$ V (Supplementary Figs. 5d and 5e). This indicates a higher energy barrier in the nonlocal transport and implies an anomalous origin of the nonlocal signal.

To determine the origin of the observed $R_{NL}$, we investigate the scaling relation between $R_{NL}$ and $R_L$ as functions of $V_g$ at different temperatures for both sample A (Fig. 2) and sample B (Supplementary Fig. 6). For a fixed $V_g$, both $R_L$ and $R_{NL}$ increase when the temperature is lowered. In sample A, two regimes with distinct scaling behaviors become clearly visible in Fig. 2c, the logarithmic plot of $R_L$ and $R_{NL}$ at different temperatures. Above 160 K, the slopes of the ln$R_{NL}$ versus ln$R_L$ curves are 1, indicating that the nonlocal signal is proportional to the classical contribution. Below 160 K, the slopes turn to 3 in the low electron density regime ($R_L \approx 10^8$–$10^9$ Ω), which amounts to $R_{NL} \propto R_L^3$. Such a cubic scaling is reminiscent of the nonlocal charge transport mediated by spin diffusion in the spin Hall effect[32]. As introduced above and calculated later (Fig. 4), the massive Dirac band structure produces large valley Hall conductivity $\sigma_{xy}^V$ (ref. 13,14) for monolayer MoS$_2$. Thus, it is natural to attribute the observed nonlocal signal to the VHE, and the VHE mediated $R_{NL}$ follows[32]

$$R_{NL} = \frac{1}{2}\left(\frac{\sigma_{xy}^V}{\sigma_{xx}}\right)^2 \frac{W}{\sigma_{xx} l_V} e^{-\frac{L}{l_V}} \propto (\sigma_{xy}^V)^2 R_L^3. \qquad (1)$$

Here $l_V$ is the valley diffusion length (or inter-valley scattering length), and $\sigma_{xx}$ and $R_L$ have the simple relation of $\sigma_{xx} = \frac{L}{R_L W}$.

The $R_{NL}$ and $R_L$ data measured at different temperatures for the case of $V_g = -60$ V are plotted in Fig. 2d. The cubic law is not applicable above 160 K, probably due to the suppression of valley Hall conductivity at high temperatures. This is consistent with previous experiments in bilayer graphene[9,11]. Below 160 K, Eq. (1) can be employed to estimate $l_V$. For the case of intermediate inter-valley scattering and edge roughness, $l_V \sim 0.36$ μm if we assume $\sigma_{xy}^V \sim 1$ e$^2$/h. In the limit of strong inter-valley scattering and edge roughness, $l_V \sim 0.43$ μm if we assume $\sigma_{xy}^V \sim 0.1$ e$^2$/h. These values of $l_V$ are comparable to those obtained in graphene systems[8-12,15,16].

We further investigated the length dependence of the nonlocal valley transport. Apart from sample A ($L = 3.6$ μm) and sample B ($L = 6$ μm), two more samples ($L = 11$ μm and 16 μm) are investigated (Supplementary Fig. 6). The semilog plot of $R_{NL}$ at $n = 4\times10^{10}$ cm$^{-2}$ (extracted from the Hall measurement, see Supplementary Fig. 8) versus the sample length yields an estimation of $l_V \sim 1$ μm (Fig. 2g). This value is very close to W and much larger than the electron mean free path $l_m \sim 20$ nm estimated from the sample mobility $\mu$ for the range of n where the cubic scaling appears. Nevertheless, these estimations based on the observed nonlocal signals are suggestive of $l_V$ in the order of micron. In sample B, the cubic scaling remains even at room temperature, attributed to the dominant valence-band contribution to $\sigma_{xy}^V$ and particularly the large intrinsic bandgap that is impossible for graphene systems.

**Nonlocal transport in bilayer and trilayer MoS$_2$.**

For bilayer MoS$_2$, the measured $R_L$ and $R_{NL}$ as functions of $V_g$ at different temperatures are plotted in Figs. 3a–3b. As the carrier density increases, $R_L$ and $R_{NL}$ decrease in a similar fashion in the temperature range of 5–50 K. This yields a linear scaling behavior between $R_L$ and $R_{NL}$, as analyzed in Figs. 3c–3d, and no cubic scaling is detected. We note that extrinsic $P$ symmetry breaking can be introduced into atomically thin bilayers via external gating, as achieved in bilayer graphene[9,10], and that detecting a nonlocal signal in gated bilayer graphene requires a threshold gating strength[9,10]. In our devices, however, $V_g$ is too low to reach the threshold estimated by an recent optical experiment[16]; the estimated potential difference between the top and bottom layers is ~ 9.2 meV at $V_g = -60$ V[16]. This weak symmetry breaking produces little change in the total Berry curvature as compared with the pristine case (Supplementary Fig. 7), given the facts that the induced potential is much smaller than the bandgap and that the valence-band contribution to $\sigma_{xy}^V$ is dominant. In light of this analysis, the gating-induced $P$ symmetry breaking is negligible in our bilayer MoS$_2$. Therefore, we conclude that the absence of cubic scaling in bilayer MoS$_2$ indicates the crucial role of strong $P$ symmetry breaking in generating VHE.

This key conclusion can be immediately tested in thicker MoS$_2$ samples. Given that $P$ symmetry is broken (respected) in pristine odd-layer (even-layer) MoS$_2$, one might wonder whether the intrinsic VHE and its cubic scaling could be detected in trilayer MoS$_2$. Figs. 3e and 3f display our $R_L$ and $R_{NL}$ data measured in trilayer MoS$_2$ as functions of $V_g$ at different temperatures. Evidently, the measured $R_{NL}$ rapidly decreases as $V_g$ increases in the narrow range of $-20$ V $< V_g < -18.4$ V, which is reminiscent of the behavior of $R_{NL}$ in our monolayer devices in the low density regime. Similar to the monolayer case, the logarithmic plots of $R_L$ and $R_{NL}$ in Fig. 3g exhibit clear changes in slop from 1 to 3 near $V_g$ = −18.4 V, further confirming the observation of the nonlocal signal of VHE in trilayer $MoS_2$. To illustrate the temperature dependence, Fig. 3h plots the scaling relation between $R_L$ and $R_{NL}$ at different temperatures for the case of $V_g$ = −20 V. Again, there is a clear change in slop from 1 to 3 near 30 K. Moreover, the valley diffusion length can be extracted based on Fig. 3h and Eq. (1). We obtain $l_V \sim$ 0.5 μm and ~1 μm, respectively, for the aforementioned two limits $\sigma_{xy}^V \sim 1$ $e^2/h$ and ~0.1 $e^2/h$. Both the observed amplitude of nonlocal signal and the estimated valley diffusion length in the trilayer $MoS_2$ devices are comparable to those in the monolayer case. In addition to the crucial role of $P$ symmetry breaking, these observations are suggestive of a universal physical origin of VHEs in odd-layer TMDCs.

**Discussion**

To better understand the thickness dependent observations, we calculate[13,14] the electronic band structures and Berry curvatures for monolayer, bilayer, and trilayer $MoS_2$. The band structures in Figs. 4a-4c are indeed thickness dependent. In particular, the conduction-band minima lie at the K-valleys for the monolayer, whereas they shift to the Q-valleys for the bilayer and trilayer. Given the low electron densities in our samples (~3×10$^{10}$ cm$^{-2}$ in monolayers and ~1×10$^{11}$ cm$^{-2}$ in bilayers and trilayers), the Fermi levels only cross the lowest sub-bands, as indicated by the green lines in Figs. 4a-4c. As bilayer $MoS_2$ has a restored $P$ symmetry that is intrinsically broken in odd-layer $MoS_2$, the sub-bands are spin degenerate in the bilayer yet spin split in the monolayer and trilayer. With the band structures, we further derive the Berry curvatures that can drive the VHEs. Berry curvature vanishes if both $P$ and $T$ are present. As plotted in Figs. 4d-4f, our calculations reveal that the curvatures are indeed trivial in the bilayer yet substantial in the monolayer and trilayer. This explains the reason why no cubic scaling is observed in bilayer $MoS_2$ and highlights the role of $P$ symmetry breaking in producing VHEs.

It is puzzling to understand and compare the nonlocal signals of VHEs in monolayer and trilayer $MoS_2$. Similar cubic scaling behaviors and their transitions to linear ones above the critical densities or temperatures are observed in both cases. However, the conduction-band Berry curvatures (the difference between the blue and orange curves in Fig. 4d and Fig. 4f) are considerably large in the monolayer K-valleys yet negligibly small in the trilayer Q-valleys. This implies that the geometric VHE requiring finite doping[3] might not be the origin[9], which is further evidenced by the fact that the cubic scaling behaviors weaken rapidly with increasing the electron densities.

On the other hand, these facts appear to be in harmony with the topological VHE that arises from the valence-band contributions[4-7]. Particularly, the monolayer and trilayer share almost identical valence-band Berry curvatures (the orange curves in Fig. 4d and Fig. 4f) and valley Hall conductivities, due to the extremely weak interlayer couplings. Recently, *nearly* quantized edge transports have been observed along the designed or selected domain walls in graphene systems[11,12] and even in artificial crystals[17]. In our case, the roughness of natural edges can cause *edge* inter-valley scattering[6] and passivate the edge states[33]. The electrons in the passivated edge states are likely responsible for the nonlocal signals of VHE; these edge states are partially gapped and well spread into the interiors, leading to bulk instead of edge valley transport.

The pronounced nonlocal signals are observed in our $MoS_2$ samples with length up to 16 μm and at temperature up to 300 K. The valley diffusion lengths are also estimated to be in the order of micron. The low carrier concentration ensures the low possibility of *bulk* inter-valley scattering and maintains a long valley diffusion length. In addition, the mirror and *T* symmetries lock the spin and valley indices of the lowest sub-bands, preventing *bulk* inter-valley scattering via spin conservation. Our observed intrinsic VHEs and their long valley diffusion lengths are promising for realizing room-temperature low-dissipation valleytronics. To better elucidate the outstanding problems of both geometric[3] and topological[4-7] VHEs, our observations and analyses call for future efforts, particularly complementary experiments in p-type TMDCs.

**Methods**

(a) Van der Waals structures

MoS$_2$ bulk crystals are bought from 2Dsemiconductors (website: http://www.2dsemiconductors.com/), and the h-BN sources (grade A1) were bought from HQgraphene (website: http://www.hqgraphene.com/). To fabricate van der Waals heterostructures, a selected MoS$_2$ sample is picked up from the SiO$_2$/Si substrate by a thin h-BN flake (5-15 nm thick) on PMMA (950 A7, 500 nm) via van der Waals interactions. The h-BN/MoS$_2$ flake is then transferred onto a fresh thick h-BN flake lying on another SiO$_2$/Si substrate, to form a BN-MoS$_2$-BN heterostructure (step 1 in Supplementary Fig. 1).

(b) Selective etching process

A hard mask is patterned on the heterostructure by the standard e-beam lithography technique using PMMA (step 2 in Supplementary Fig. 1). The exposed top BN layer and MoS$_2$ are then etched via reactive ion etching (RIE), forming a Hall bar geometry (steps 3 & 4 in Supplementary Fig. 1). Then a second-round e-beam lithography and RIE is carried out to expose the MoS$_2$ layer (steps 5 & 6 in Supplementary Fig. 1). The electrodes are then patterned by a third-round e-beam lithography followed by a standard e-beam evaporation (steps 7 & 8 in Supplementary Fig. 1). To access the conduction band edges of MoS$_2$, we choose Titanium as the contact metal, as the work function of Titanium (~4.3 eV) matches the band-edge energy of MoS$_2$ (~4.0 to 4.4 eV depending on the layer numbers).

(c) Electronic measurement

The *I–V* curves are measured by Keithley 6430. Other transport measurements are carried out by using: (i) low-frequency lock-in technique (SR 830 with SR550 as the preamplifier and DS 360 as the function generator, or (ii) Keithley 6430 source meter (> $10^{16}$ Ω input resistance on voltage measurements). The cryogenic system provides stable temperatures ranging from 1.8 to 300 K. A detailed discussion of the nonlocal measurement is presented in Supplementary Fig. 4.

**Acknowledgments**

We thank Wing Ki Wong for the assistance in the sample preparation process. We acknowledge the financial support from the Research Grants Council of Hong Kong (Project Nos. 16302215, HKU9/CRF/13G, 604112 and N_HKUST613/12) and the UT-Dallas Research Enhancement Funds (toward F.Z.). We are grateful for the technical support of the Raith–HKUST Nanotechnology Laboratory at MCPF. F.Z. is grateful to Di Xiao and Allan MacDonald for valuable discussions.

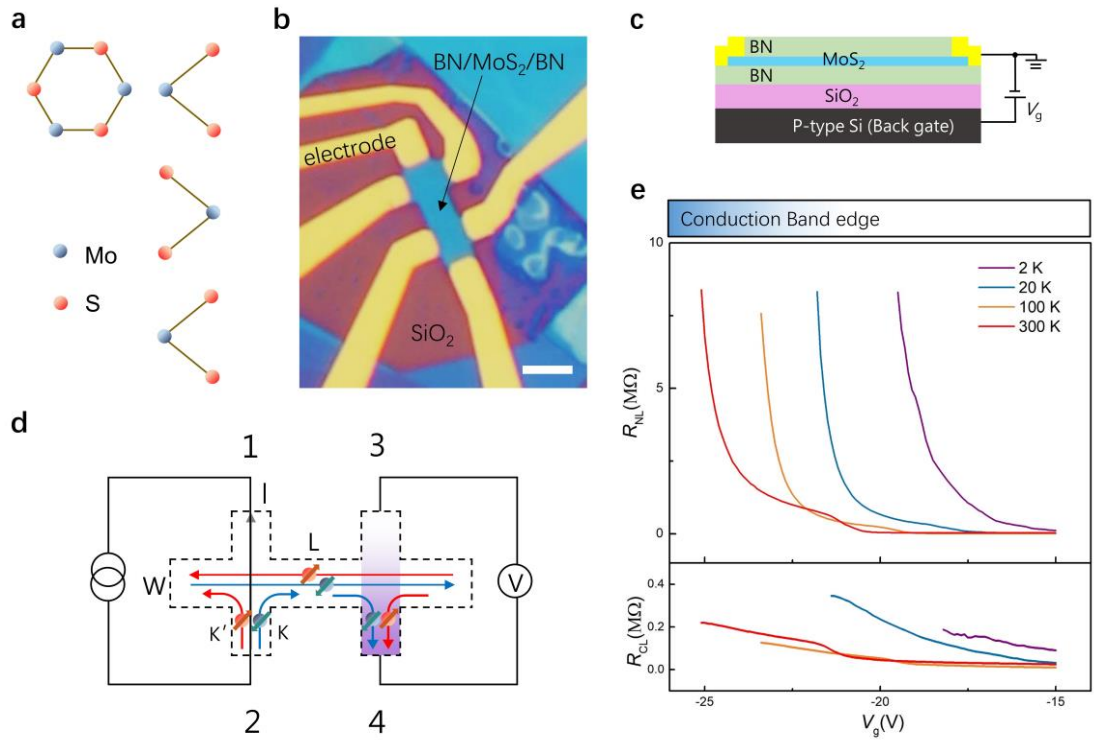

**Figure 1 | Valley Hall transport induced nonlocal resistance in a monolayer MoS₂ field-effect transistor. a**, Crystal structure of an odd/even-layer 2H-MoS$_2$ is inversion asymmetric/symmetric. **b**, Optical image of a typical monolayer MoS$_2$ device. A MoS$_2$ Hall bar is sandwiched between the top and bottom h-BN flakes. Scale bar: 3 μm. **c**, Schematic of the h-BN encapsulated MoS$_2$ field-effect transistor. **d**, Schematic of the nonlocal resistance measurement and the VHE-mediated nonlocal transport. The applied charge current in the left circuit generates a pure valley current in the transverse direction via a VHE. This valley current induces opposite chemical potentials gradients for the two valleys over the inter-valley scattering length, which, in turn, generates a voltage drop measured by probes 3 and 4 in the right circuit via an inverse VHE. **e**, Nonlocal resistance $R_{NL}$ (upper panel) and the classical ohmic contribution $R_{CL}$ (lower panel) as functions of gate voltage $V_g$ at varied temperatures.

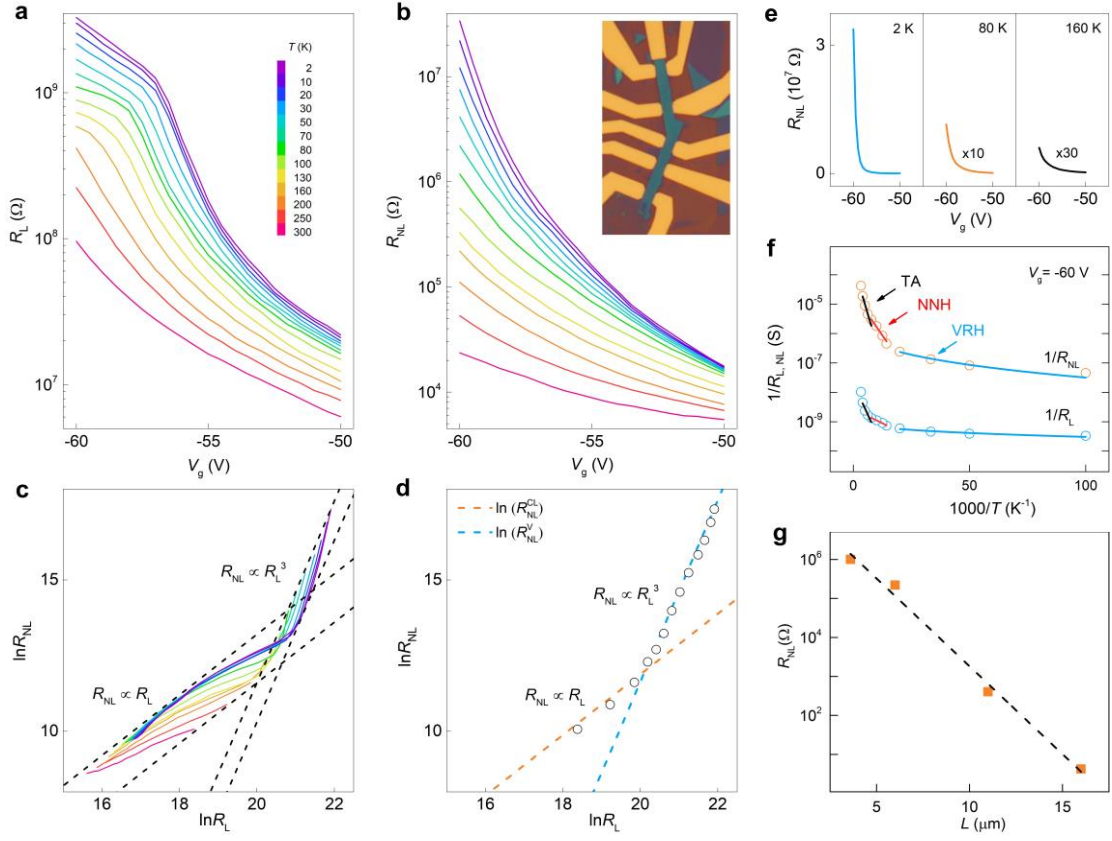

**Figure 2 | Local and nonlocal resistances of monolayer MoS$_2$. a**, **b**, Semilog plots of $R_L$ and $R_{NL}$ as a function of $V_g$ measured at varied temperatures. Inset of **b**: optical micrograph of our typical h-BN/MoS$_2$/h-BN device with multi-terminal Hall Bar configurations. **c**, Scaling relation between ln $R_L$ and ln $R_{NL}$ at temperatures ranging from 300 K to 2 K. When the electron density is relatively high, i.e., $R_L$ and $R_{NL}$ are small, $R_{NL}$ is linearly proportional to $R_L$. Below 200 K, $R_{NL}$ and $R_L$ follow a cubic relationship at low electron densities, indicating that the contribution from valley Hall transport dominates. The crossover from linear to cubic scaling is observed around the critical density of $n_c = 5 \times 10^{10}$ cm$^{-2}$, with gate voltage $V_g = -57$ V. **d**, Crossover phenomenon by considering classical diffusion ($R_{NL} \propto R_L$) and valley Hall transport ($R_{NL} \propto R_L^3$). The experimental data (black circles, $V_g = -60$ V) clearly show two different regimes which are fitted by two linear curves (orange dashed line with slope 1 and blue dashed line with slope 3). **e**, $R_{NL}$ plotted as a function of $V_g$ at

low temperatures. The ohmic contribution, calculated according to $R_L$ and device geometry, is deducted from the measured $R_{NL}$ at different temperatures. **f,** $1/R_{NL}$ (orange circles) and $1/R_L$ (blue circles) in log scale plotted as functions of $1/T$ at $V_g = -60$ V. Three distinct transport regimes were observed: the thermal activation (TA) transport, nearest neighbor hopping (NNH) transport, and the variable range hopping (VRH) transport. **g,** Semilog plot of $R_{NL}$ as a function of L at $n = 4 \times 10^{10}$ cm$^{-2}$ (orange squares). Nonlocal signal decays exponentially with increasing L. The dashed line yields a valley diffusion length of ~ 1 μm.

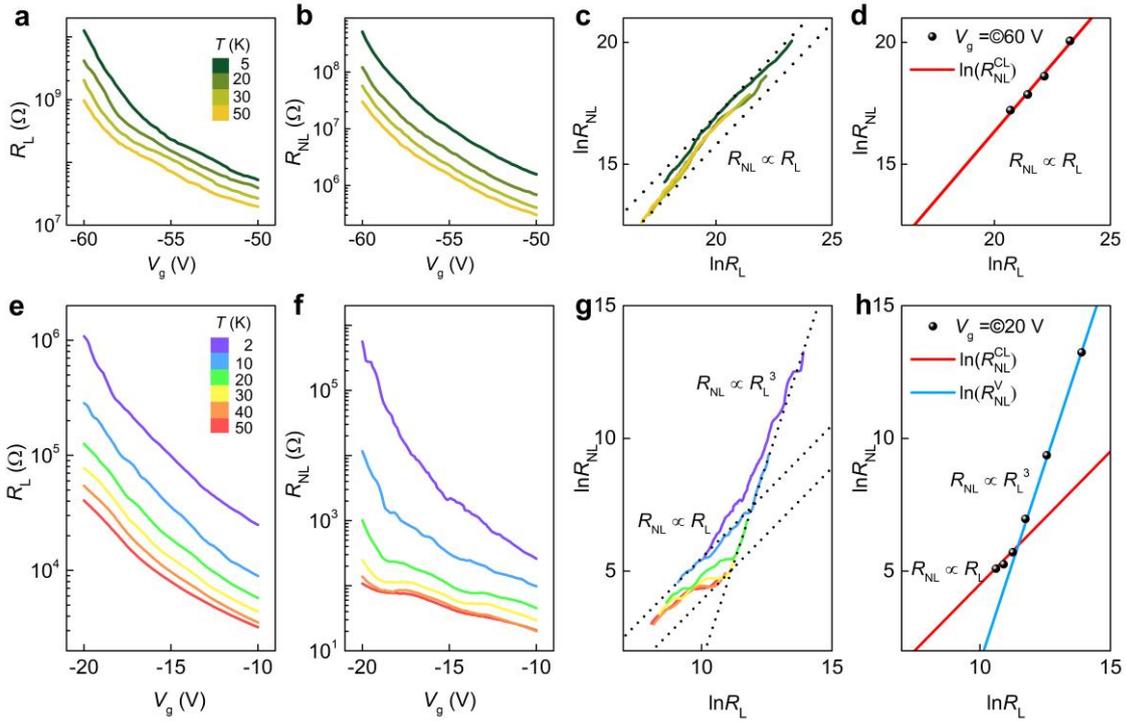

**Figure 3 | Local and nonlocal resistances of bilayer and trilayer MoS₂. a,b,e,f,** Gate-dependence of $R_L$ and $R_{NL}$ at different temperatures in bilayer (**a,b**) and trilayer (**e,f**) samples. **c,g,** Scaling relation between $\ln R_L$ and $\ln R_{NL}$ is obtained at different temperatures in bilayer (**c**) and trilayer (**g**) samples. For the trilayer case, $R_{NL}$ scales linearly with $R_L$ in the high electron density regime, whereas the cubic scaling law $R_{NL} \propto R_L^3$ is observed in the low electron density regime ($n_c = 1.1 \times 10^{12}$ cm$^{-2}$ or $V_g = -18.4$V). **d,** $\ln R_L$ v.s. $\ln R_{NL}$ for bilayer MoS₂. In the full range of gate voltages, $R_{NL}$ scales linearly with $R_L$, and the experimental data (black dots, $V_g = -60$ V) is fitted by a linear curve (red solid line). **h,** $\ln R_L$ v.s. $\ln R_{NL}$ for trilayer MoS₂. The experimental data (black dots, $V_g = -20$ V) clearly show two different regimes which are fitted by two linear curves (red solid line with slope 1 and blue solid line with slope 3). Evidently, a crossover exists from linear ($R_{NL} \propto R_L$) to cubic scaling behaviors ($R_{NL} \propto R_L^3$).

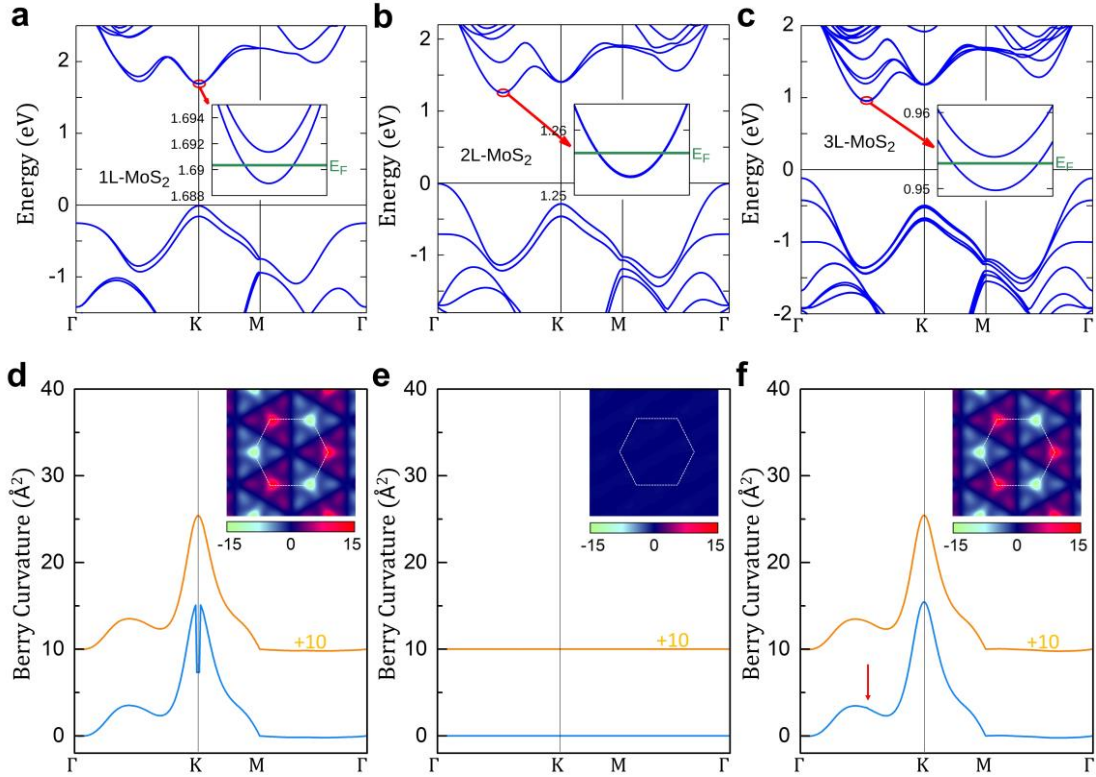

**Figure 4 | Calculated band structures and Berry curvatures of monolayer, bilayer, and trilayer MoS$_2$. a-c**, Band structure of (**a**) monolayer, (**b**) bilayer, and (**c**) trilayer MoS$_2$. The conduction band edges lie at the K-valleys in the monolayer but at the Q-valleys in the bilayer and trilayer. Insets of **a-c**: The Fermi levels only cross the lowest sub-bands, which are spin degenerate in (**b**) but spin split in **a** and **c**. **d-f**, Berry curvatures of (**d**) monolayer, (**e**) bilayer, and (**f**) trilayer MoS$_2$. The blue curves are the total curvatures of all occupied states below the Fermi levels, whereas the orange curves are the total curvatures of all valence-band states. The red arrow in **f** points out a tiny bump at a Q valley. Insets of **d-f**, 2D mapping of Berry curvatures in the 2D Brillouin zone (white dashed lines).